\documentclass[onecolumn,tighten]{aastex631}

\shorttitle{A Stochastic-Variability Transition in 1ES 1927+654}
\shortauthors{Dong \& Yan}
\graphicspath{{figures/}}
\usepackage{amsmath}
\usepackage{multirow}
\usepackage{tipa}

\usepackage{rotating}

\begin{document}
\title{Gaussian-process evidence for a stochastic-variability transition in the recovering corona of 1ES 1927+654}
\author[0009-0008-0162-9771]{Lijuan Dong}
\affiliation{Department of Astronomy, School of Physics and Astronomy, Yunnan University, Kunming, Yunnan, 650091, People’s Republic of China}
\author[0000-0003-4895-1406]{Dahai Yan}
\correspondingauthor{Dahai Yan}{\email{yandahai@ynu.edu.cn}}
\affiliation{Department of Astronomy, School of Physics and Astronomy, Yunnan University, Kunming, Yunnan, 650091, People’s Republic of China}

\begin{abstract}
We investigate the stochastic X-ray variability of the changing-look active galactic nucleus 1ES 1927+654 during its 2018--2024 evolution, focusing on the recovery of the X-ray corona after its 2018 collapse. Using XMM-Newton EPIC-pn light curves in the 0.3--2.0 keV and 2.0--10.0 keV bands, we model the variability with Gaussian process (GP) covariance components including Mat\'ern-3/2, damped-random-walk (DRW), stochastically driven damped simple-harmonic-oscillator (SHO), and white-noise terms. Bayesian model comparison reveals an X-ray stochastic-variability transition during the changing-look recovery phase. In the 2019 May 5 observation, the preferred covariance changes from a Mat\'ern-3/2-like state to a DRW-like state within a single continuous exposure. 
A phenomenological gated-kernel estimate localizes this transition sharply in the hard band at $t_c\simeq23.5~{\rm ks}$, while the soft band shows the same qualitative change over a broader interval. This transition occurs after the X-ray corona had reappeared but before the later pronounced hardening and brightening of the coronal emission, suggesting an early timing-domain signature of disk--corona reconfiguration. Phenomenologically, the dominant variability evolves from a smoother, finite-memory correlated process to a rougher, shorter-memory red-noise process. In the later 2022--2024 observations, SHO-like components associated with the known millihertz QPO show increasing characteristic frequency and quality factor, indicating a faster and more coherent oscillatory component during the QPO-plus-jet phase. GP-based time-domain inference therefore provides a sensitive probe of stochastic-variability changes in recovering AGN coronae.

\end{abstract}


\section{Introduction} \label{sec:intro}
1ES~1927+654 is a changing-look active galactic nucleus (AGN) at a redshift of $z = 0.019422$ \citep{2019ApJ...883...94T}. 
The source has undergone
dramatic changes in its emission properties. Its central black hole has a mass of $M_{\rm BH}=1.38_{-0.66}^{+1.25}\times10^6\,M_{\odot}$ \citep{2022ApJ...931....5L}. Before 2017, the source was classified as a true type-II AGN, because it did not show broad emission lines and strong obscuration \citep{2003A&A...397..557B,2019ApJ...883...94T}. In late 2017, 1ES 1927+654 brightened by nearly two orders of magnitude in the optical and UV, and broad Balmer lines appeared by March 2018, marking a transition from a Seyfert 2 to a Seyfert 1 state \citep{2019ApJ...883...94T,2022ApJ...934...35M}. This event has been interpreted as a TDE-like outburst in an existing AGN system \citep{2020ApJ...898L...1R,2023MNRAS.526.2331C}. 
% The X-ray evolution was more dramatic. 
After May 2018, the X-ray flux declined rapidly, and by August 2018 the corona had almost disappeared \citep{2020ApJ...898L...1R}. 
It reappeared later in 2018, brightened to about ten times the preflare level by November 2019, and returned close to its preflare state by May 2021 \citep{2021ApJS..255....7R,2022ApJ...931....5L,2022ApJ...934...35M}.

Previous studies have described several key aspects of this recovery process. 
Broadband spectral modeling showed that the accretion flow evolved from a super-Eddington slim disk to a thin disk dominated by radiation pressure \citep{2024ApJ...975...50L}. The hard X-ray power-law component almost disappeared after the outburst and then gradually reappeared, suggesting the destruction and later recreation of the corona \citep{2020ApJ...898L...1R}. Since May 2022, the source has shown a renewed rise in soft X-ray emission, without a comparable optical or UV flare \citep{2023ApJ...955....3G}. XMM-Newton observations in July 2022 revealed a millihertz QPO in the X-ray band \citep{2025Natur.638..370M}. 
The QPO frequency later increased and then approached a stable value of about 2.5 mHz \citep{2025Natur.638..370M,2026arXiv260416688M}. This QPO has been interpreted as an oscillation of a compact Comptonizing corona close to the black hole \citep{2025Natur.638..370M,2025arXiv251104264S}. Radio observations also revealed a rapid flare and a bipolar jet-like outflow, suggesting a possible link between the coronal recovery, the QPO phase, and jet formation \citep{2025ApJ...979L...2M,2025ApJ...981..125L}. However, it remains unclear whether the coronal recovery left a detectable imprint on the stochastic structure of the X-ray light curves themselves.

Gaussian-process (GP) methods provide a flexible framework for modeling stochastic and quasi-periodic variability in astronomical time series. They have been widely applied to time-series modeling, AGN variability,
periodicity searches, and QPO studies
\citep{2012RSPTA.37110550R,2018ApJ...869..178T,2021ApJ...914..144G,2021ApJ...907..105Y,2022MNRAS.513.2841C,2022ApJ...936...17H,
2025MNRAS.537.3210G}. In AGNs, GP methods have been used to search for $\gamma$-ray QPOs \citep{2021ApJ...919...58Z}, measure multi-band damping timescales and
examine their connection with black-hole mass
\citep{2022ApJ...930..157Z,2023ApJ...944..103Z}, characterize stochastic variability during major GeV flares in blazars
\citep{2025ApJ...988..206Z}, and decompose QPO and 
stochastic-variability components \citep{2025MNRAS.540.3790Z}.
GP-based approaches have also been recently applied to accreting black-hole X-ray binaries
\citep{2025A&A...703A.134W,2026arXiv260401901D}.

% GP methods provide a flexible way to test for such stochastic-variability changes through the covariance structure of the light curve. In AGN studies, they have been used to search for $\gamma$-ray QPOs \citep{2021ApJ...919...58Z}, to measure multi-band damping timescales and examine their connection with black-hole mass \citep{2022ApJ...930..157Z,2023ApJ...944..103Z}, to decompose the PG 1553+113 $\gamma$-ray light curve into QPO and stochastic-background components \citep{2025MNRAS.540.3790Z}, and to characterize stochastic variability during major GeV flares in blazars \citep{2025ApJ...988..206Z}. GP-based approaches have also been applied to accreting black hole X-ray binaries, where they help model QPOs together with broadband stochastic variability \citep{2025A&A...703A.134W,2026arXiv260401901D}.

In this work, we apply the \textsc{celerite2} GP framework \citep{celerite1,celerite2} to the XMM-Newton EPIC-pn light curves of 1ES 1927+654 in the 0.3--2.0 keV and 2.0--10.0 keV bands. 
We identify an X-ray stochastic-variability transition during the changing-look recovery phase: in the 2019 May 5 observation PN\_0843270101, the preferred covariance changes from a Mat\'ern-3/2-like state to a DRW-like state within a single continuous exposure. 
This transition is sharply localized in the hard band at $t_c\simeq23.5~{\rm ks}$, while the soft band shows the same qualitative change over a broader interval. We characterize this covariance-state change as a quantifiable stochastic-variability transition feature and place it in the broader 2018--2024 X-ray timing evolution of the source.
%We interpret it as an early timing-domain marker of disk--corona reconfiguration and connect it to the broader 2018--2024 evolution, including the later SHO-like components associated with the millihertz QPO.

%This Letter is organized as follows. 
%Section~\ref{sec:data and method} describes the sample and method, including the GP model and the model selection procedure. 
%Section~\ref{sec:results} presents the main results, including the parameter evolution in the two energy bands, the stochastic transition within PN\_0843270101, and the SHO parameter evolution in the 2.0--10.0 keV band. 
%Section~\ref{sec:scenario} discusses the stochastic magneto-active corona scenario, links the GP results to the coronal recovery process, and interprets the kernel transition during PN\_0843270101. 
%Section~\ref{sec:conclusions} summarizes the main conclusions.

\section{Data and Time-Domain Modeling}
\label{sec:data and method}

\subsection{XMM-Newton observations and energy bands}

We use the published XMM-Newton EPIC-pn light curves of 1ES 1927+654 from the X-ray timing analysis of \citet{2025Natur.638..370M}. 
EPIC-pn is adopted because of its high time resolution and sensitivity to rapid X-ray variability \citep{2001A&A...365L...1J}. 
We analyze the 0.3--2.0 keV and 2.0--10.0 keV bands, which include 15 observations from 2018 May to 2024 March and 14 observations from 2018 November to 2024 March, respectively. 
These data span the changing-look recovery phase, the later millihertz-QPO phase, and the jet phase.

\subsection{Gaussian-process representation of stochastic variability states}

We use GP models to characterize the stochastic X-ray variability of 1ES 1927+654.
A GP defines a probability distribution over functions that can represent possible light curves. It is specified by a mean function and a covariance function \citep{2006gpml.book.....R,2023ARA&A..61..329A}.
In this work, the covariance function is used as a phenomenological description of the dominant variability state.
We therefore do not assign a unique physical mechanism to any individual kernel.
Instead, we use Bayesian model comparison to determine which stochastic representation best describes each light curve or time interval.

We consider four basic covariance components: a damped random walk (DRW), a Mat\'ern-3/2 process, a stochastically driven damped simple harmonic oscillator (SHO), and an additional white-noise (Wn) term.
The DRW, Mat\'ern-3/2, and SHO components are adopted because they admit simple stochastic differential-equation representations \citep[e.g.,][]{2022ApJ...930..157Z}, allowing their covariance properties to be related to characteristic dynamical behavior. This facilitates a dynamical interpretation of the inferred variability
states, while the Wn term accounts for additional uncorrelated scatter.

The DRW process is the stationary solution of a first-order stochastic relaxation equation, with the covariance kernel given by
\begin{equation}
k_{\rm DRW}(\tau)=a\exp(-c\tau),
\end{equation}
where $\tau$ is the time lag, $a$ is the amplitude, and $c$ is the damping rate.
The corresponding characteristic timescale is $\tau_{\rm DRW}=1/c$ \citep{1992ApJ...398..169R}.

The Mat\'ern-3/2 process admits a linear stochastic differential-equation representation and exhibits smoother short-lag behavior than the DRW process. Its covariance kernel is given by \citep{2006gpml.book.....R},
\begin{equation}
k_{\rm M32}(\tau)=\sigma^2\left(1+\frac{\sqrt{3}\tau}{\rho}\right)
\exp\left(-\frac{\sqrt{3}\tau}{\rho}\right),
\end{equation}
where $\sigma$ is the amplitude and $\rho$ is the characteristic correlation timescale.

The SHO component describes damped quasi-periodic variability as the
stationary response of a damped harmonic oscillator to stochastic driving.
In the \textsc{celerite2} formulation, the process is governed by
\citep{celerite2}
\begin{equation}
\left[
\frac{\mathrm{d}^{2}}{\mathrm{d}t^{2}}
+
\frac{\omega_{0}}{Q}\frac{\mathrm{d}}{\mathrm{d}t}
+
\omega_{0}^{2}
\right]y(t)
=
\epsilon(t),
\end{equation}
where $\epsilon(t)$ denotes the stochastic driving term,
$\omega_{0}=2\pi\nu_{0}$ is the undamped natural angular frequency, $\nu_{0}$ is the corresponding characteristic oscillation frequency, and $Q$ is the quality factor that controls the damping and coherence of the oscillation. 
In the \textsc{celerite2} parameterization, $S_0$ sets the normalization of the SHO component.
The SHO kernel is derived from the linear dynamics of a stochastically driven, damped harmonic oscillator. Assuming Gaussian white-noise forcing and stationarity, the analytical power spectral density of the output process is first obtained from the oscillator’s frequency-response function. The corresponding stationary covariance function is then derived by inverse Fourier transformation according to the Wiener–Khinchin theorem. Finally, this covariance is expressed as a combination of exponentially damped sine and cosine terms and matched to the standard celerite kernel form, enabling efficient likelihood evaluation through its semiseparable matrix representation \citep{celerite1}.

The Wn component accounts for an extra uncorrelated scatter beyond the reported statistical errors.
It contributes only to the diagonal elements of the covariance matrix,
\begin{equation}
k_{\rm Wn}(t_n,t_m)=\sigma_{\rm n}^{2}\delta_{nm},
\end{equation}
where $\sigma_{\rm n}^{2}$ is the additional white-noise variance and $\delta_{nm}$ is the Kronecker delta.

These covariance components are implemented using the structured
kernel representation of \textsc{celerite2}, which enables efficient
factorization of the GP covariance matrix and likelihood evaluation
without constructing a dense matrix
\citep{celerite1,celerite2}.
For observations $\boldsymbol{y}$ with mean model $\boldsymbol{\mu}$,
the GP log-likelihood is
\begin{equation}
\ln \mathcal{L}
=
-\frac{1}{2}
(\boldsymbol{y}-\boldsymbol{\mu})^{\rm T}
\mathbf{C}^{-1}
(\boldsymbol{y}-\boldsymbol{\mu})
-\frac{1}{2}\ln|\mathbf{C}|
-\frac{N}{2}\ln(2\pi),
\end{equation}
where $\mathbf{C}$ is the covariance matrix including the GP covariance,
measurement uncertainties, and, where required, the additional Wn term.

For each light curve, we construct candidate covariance models from these components, either as single-component models or as sums of components.

\subsection{Bayesian model comparison with \texttt{dynesty}}
\label{Bayesian model}

We use \texttt{dynesty} to perform Bayesian parameter inference and model
comparison through nested sampling \citep{2020MNRAS.493.3132S}.
For a light-curve data set $D$ and a candidate GP model $M$, the posterior
distribution of the model parameters $\theta$ is

\begin{equation}
p(\theta \mid D,M)
=
\frac{\mathcal{L}(D\mid\theta,M)\,\pi(\theta\mid M)}
{Z_M},
\qquad
Z_M
=
\int
\mathcal{L}(D\mid\theta,M)\,
\pi(\theta\mid M)\,d\theta .
\end{equation}

Here $\mathcal{L}$ is the GP likelihood, $\pi$ is the prior, and $Z_M$ is
the Bayesian evidence (marginal likelihood). The posterior samples
constrain the GP parameters, while $Z_M$ is used to compare competing
covariance models.

Nested sampling evaluates the Bayesian evidence by transforming the
multi-dimensional parameter integral into a one-dimensional integral
over the enclosed prior volume,
\begin{equation}
X(\lambda)
=
\int_{\mathcal{L}(\theta)>\lambda}\pi(\theta)\,d\theta,
\qquad
Z
=
\int_0^1 \mathcal{L}(X)\,dX .
\end{equation}
As the likelihood threshold increases, the enclosed prior volume
decreases from unity toward zero. \texttt{dynesty} explores these nested
likelihood contours to estimate $Z$ and generate weighted posterior
samples, which are used to constrain the GP parameters and their
correlations. The GP likelihood is evaluated using \textsc{celerite2},
and the \texttt{prior\_transform} function maps samples from the unit hypercube onto the adopted parameter priors.
The adopted prior distributions and parameter ranges for the GP parameters are summarized in Tab.~\ref{tab:priors}.

\begin{table}[htbp]
\centering
\caption{Prior distributions adopted for the GP model parameters.}
\label{tab:priors}

\renewcommand{\arraystretch}{1.6}

\begin{tabular}{lcc}
\hline
\textbf{Model} & \textbf{Parameter} & \textbf{Prior} \\
\hline

\multirow{3}{*}{SHO}
& $\ln S_{0}$        & Uniform $(-10,\,10)$ \\
& $\ln Q$            & Uniform $(-10,\,10)$   \\
& $\ln \nu_{0}$   & Uniform $(-10,\,5)$   \\

\hline

\multirow{2}{*}{DRW}
& $\ln \sigma_{\rm DRW}$ & Uniform $(-5,\,5)$  \\
& $\ln \tau_{\rm DRW}$   & Uniform $(-1,\,15)$ \\

\hline

\multirow{2}{*}{Mat\'ern-3/2}
& $\ln \sigma_{\rm M32}$ & Uniform $(-5,\,5)$  \\
& $\ln \rho_{\rm M32}$   & Uniform $(-1,\,15)$ \\

\hline

Wn
& $\ln \sigma_{n}$ & Uniform $(-5,\,5)$ \\

\hline
\end{tabular}

\end{table}

For two candidate models ($M_1$ and $M_2$), we define $\Delta \ln Z=\ln Z_1-\ln Z_2$, so that positive and negative values favor $M_1$ and $M_2$, respectively. 
The strength of the preference is assessed using the Jeffreys scale \citep{2008ConPh..49...71T}: the comparison is inconclusive for $|\Delta \ln Z|<1$, significant for $1\leq |\Delta \ln Z|<2.5$, strong for $2.5\leq |\Delta \ln Z|<5$, and decisive for $|\Delta \ln Z|\geq5$.

\section{Results}
\label{sec:results}

\subsection{Stationary GP analysis for PN\_0843270101 on 2019 May 5}
\label{sec}

\begin{figure*}[t!]
\centering
\begin{minipage}[b]{0.48\linewidth}
\centering
\includegraphics[width=\linewidth]{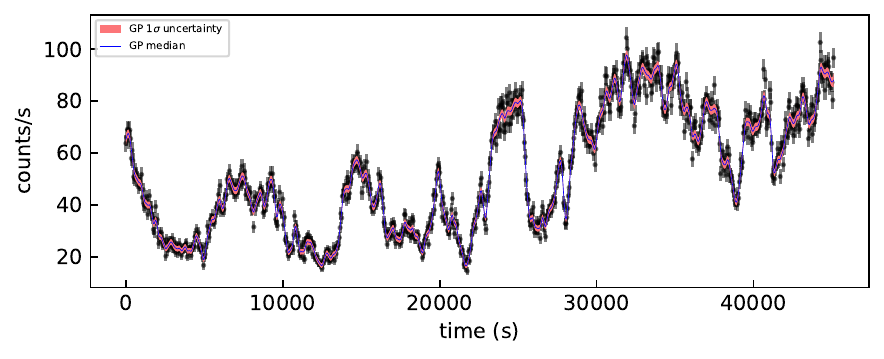}
\end{minipage}
\hfill
\begin{minipage}[b]{0.48\linewidth}
\centering
\includegraphics[width=\linewidth]{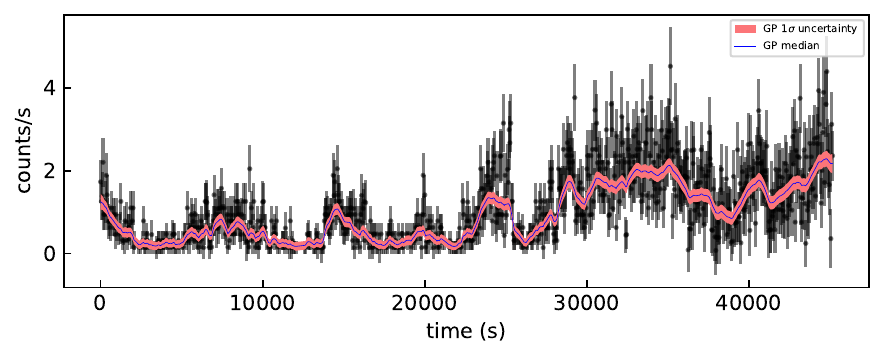}
\end{minipage}

\vspace{0.5em}

\begin{minipage}[b]{0.48\linewidth}
\centering
\includegraphics[width=\linewidth]{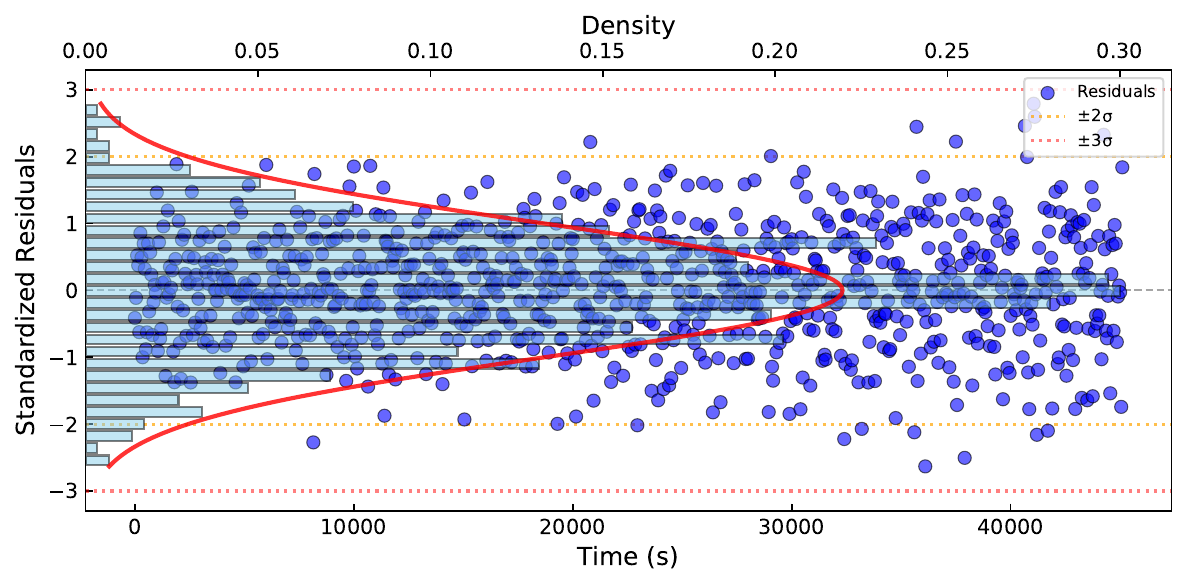}
\end{minipage}
\hfill
\begin{minipage}[b]{0.48\linewidth}
\centering
\includegraphics[width=\linewidth]{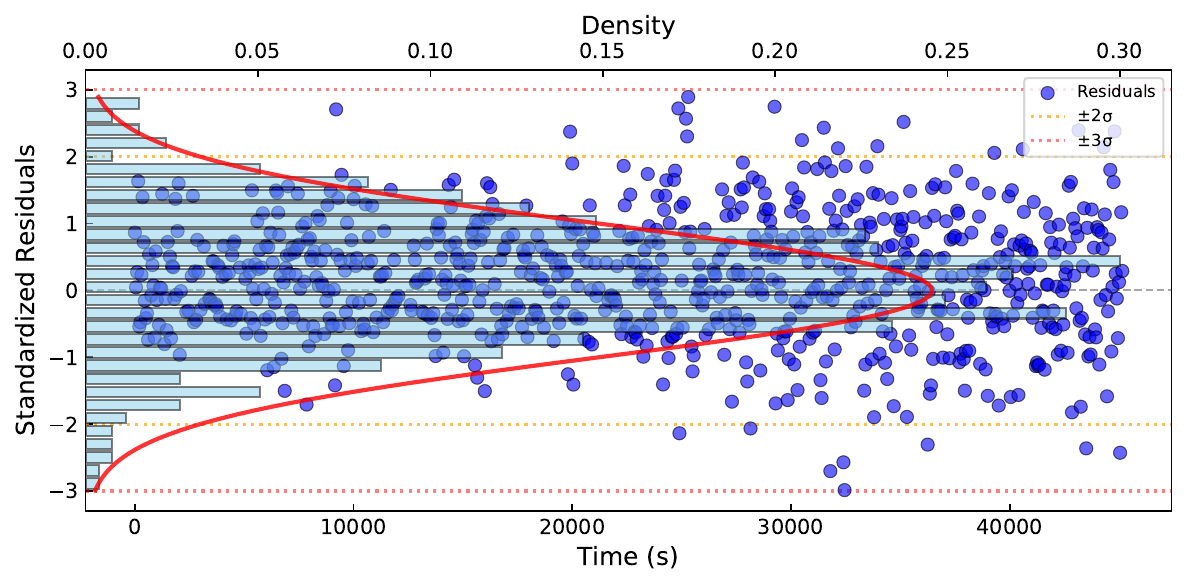}
\end{minipage}

\caption{
Single-state GP fits and residual diagnostics for PN\_0843270101.
The left column shows the 0.3--2.0 keV band.
The right column shows the 2.0--10.0 keV band.
The first row shows the light curves, the GP fit, and the $1\sigma$ uncertainty.
The second row shows the standardized residuals and their density distributions.
The red curves show the expected normal envelope.
The dashed lines mark the $\pm 2\sigma$ and $\pm 3\sigma$ levels.
}
\label{Fig1}
\end{figure*}

We first model the full exposure in each band with a single stationary GP. 
Figure~\ref{Fig1} presents the single-state GP fits and residual diagnostics for PN\_0843270101. 
The left and right columns show the 0.3--2.0 keV and 2.0--10.0 keV bands, respectively. 
From top to bottom, the panels show the light curves with the GP fit and $1\sigma$ uncertainty, the standardized residuals as a function of time with their marginal probability-density estimates.

The preferred single-state model is a Mat\'ern-3/2 kernel in the 0.3--2.0 keV band, and a DRW kernel in the 2.0--10.0 keV band. 
In both cases, the GP fit follows the broad light-curve evolution, but the standardized residuals reveal remaining time-dependent structure. 
Although the residuals are broadly centered on zero, their dispersion is not constant: they are relatively compact at early times, whereas the later part of the exposure shows larger scatter and more frequent excursions beyond the $\pm2\sigma$ range. 
This indicates that the variability is not fully captured by a stationary process with time-independent residual variance.

% This residual non-stationarity is important for interpreting the LSP. 
% The periodograms show features near $\sim 13$--$14$ ks, but localized changes in the mean level, variance, flare morphology, or stochastic variability state can be projected by the global sinusoidal basis into apparent low-frequency excess or a spurious spectral break. 
% We therefore treat these features as candidate timescales associated with the observed non-stationary light-curve evolution. 

Motivated by the residual structure, we next apply a gated-kernel timing analysis to estimate whether and when the dominant stochastic covariance state changes.

\subsection{Gated-kernel estimate of the transition time for PN\_0843270101}
\label{sec:gated_changepoint}

Stationary GP models assume that the covariance structure remains
unchanged over the analyzed interval. This assumption may become
inadequate when the stochastic variability evolves within a single
exposure, as suggested by the residual structure of PN~0843270101.
We therefore introduce a non-stationary gated-kernel model to describe
a possible transition between two covariance states
\citep{Saatci2010,2014arXiv1402.4304L}.

The model combines a pre-transition covariance
($K_{\rm pre}$) and a post-transition covariance
($K_{\rm post}$) through a smooth time-dependent gate.
To improve numerical conditioning, we first standardize the observing
times as
\begin{equation}
t_{\rm m}
=
\frac{t-\bar{t}}{\sigma_t},
\end{equation}
where $\bar{t}$ and $\sigma_t$ are the mean and standard deviation of
the observing times. The post-transition weight is defined as
\begin{equation}
q(t)
=
\left\{
1+
\exp\left[
-\left(
s[t_{\rm m}-c_{\rm m}]
+r_\theta(t_{\rm m})
\right)
\right]
\right\}^{-1},
\end{equation}
where $c_{\rm m}$ is the center of the underlying logistic transition
in standardized time and $s$ controls its sharpness.
The bounded correction term $r_\theta(t_{\rm m})$ allows small local
departures from a purely logistic transition while leaving the dominant
transition controlled by the logistic component. In our fits,
$r_\theta(t_{\rm m})$ remains small and introduces only minor
modifications to the inferred gate.

The corresponding pre- and post-transition weights are
\begin{equation}
w_{\rm pre}(t)=1-q(t),
\qquad
w_{\rm post}(t)=q(t),
\end{equation}
and the resulting non-stationary covariance is
\begin{equation}
\begin{split}
K(t,t')
={}&
w_{\rm pre}(t)w_{\rm pre}(t')
K_{\rm pre}(t,t')
\\
&+
w_{\rm post}(t)w_{\rm post}(t')
K_{\rm post}(t,t').
\end{split}
\end{equation}
This construction is positive semidefinite because it is the sum of
covariance functions multiplied by deterministic weights. When
$q(t)\simeq0$, the covariance is dominated by $K_{\rm pre}$, whereas
$q(t)\simeq1$ corresponds to the post-transition covariance
$K_{\rm post}$.

The center of the underlying logistic transition in the original time
coordinate is
\begin{equation}
t_c
=
\bar{t}+\sigma_t c_{\rm m}.
\end{equation}
For a purely logistic gate, $r_\theta=0$, $q(t_c)=1/2$, so that $t_c$
is also the midpoint of the fitted transition. When
$r_\theta(t_{\rm m})\neq0$, $t_c$ remains the parametric center of the
logistic component, while the midpoint of the full fitted gate satisfies
\begin{equation}
s(t_{\rm m}-c_{\rm m})
+
r_\theta(t_{\rm m})
=
0.
\end{equation}
The two quantities can therefore differ slightly.

We characterize the temporal extent of the transition by the effective
10--90\% width,
\begin{equation}
\Delta t_{10\text{--}90}
=
t[q(t)=0.9]
-
t[q(t)=0.1],
\end{equation}
where the two crossing times are obtained by interpolation of the
inferred $q(t)$. We report $t_c$ as the parametric transition center
and $\Delta t_{10\text{--}90}$ as the effective transition width.
Together, these quantities describe the location and temporal extent of
the inferred covariance-state transition.

The gated model is used primarily to localize a candidate transition. The gate-derived timing information is therefore followed by separate Bayesian evidence comparisons of the intervals on either side of the candidate transition
\citep{2007arXiv0710.3742P,Saatci2010}.

\begin{figure*}[t!]
  \centering
  \begin{minipage}[b]{0.48\linewidth}
    \centering
    \includegraphics[width=\linewidth]{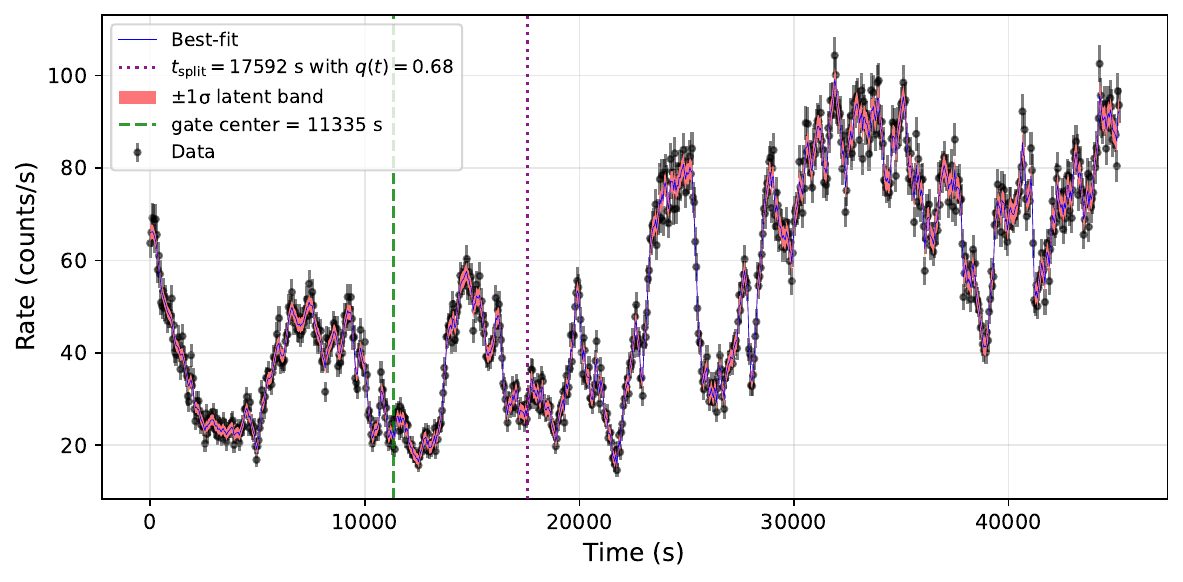}
  \end{minipage}
  \hfill
  \begin{minipage}[b]{0.48\linewidth}
    \centering
    \includegraphics[width=\linewidth]{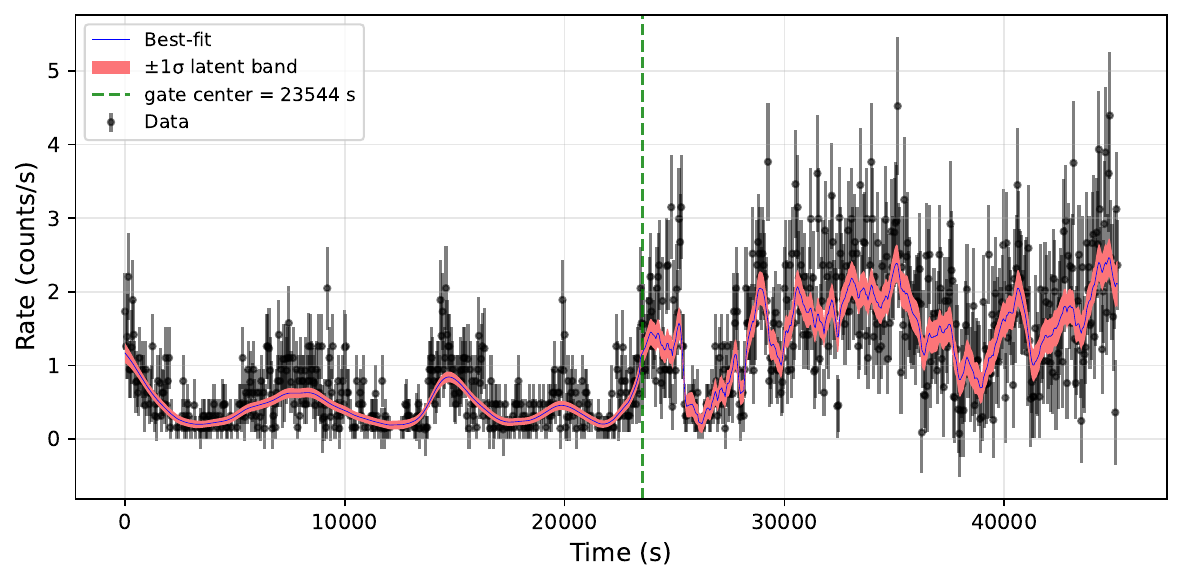}
  \end{minipage}

  \vspace{0.5em}

  \begin{minipage}[b]{0.48\linewidth}
    \centering
    \includegraphics[width=\linewidth]{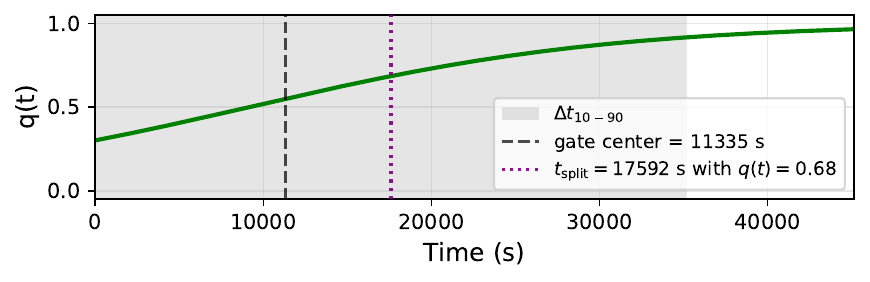}
  \end{minipage}
  \hfill
  \begin{minipage}[b]{0.48\linewidth}
    \centering
    \includegraphics[width=\linewidth]{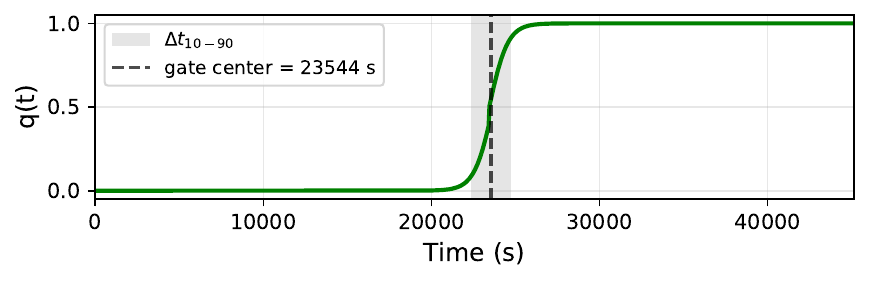}
  \end{minipage}

  \vspace{0.5em}

  \begin{minipage}[b]{0.48\linewidth}
    \centering
    \includegraphics[width=\linewidth]{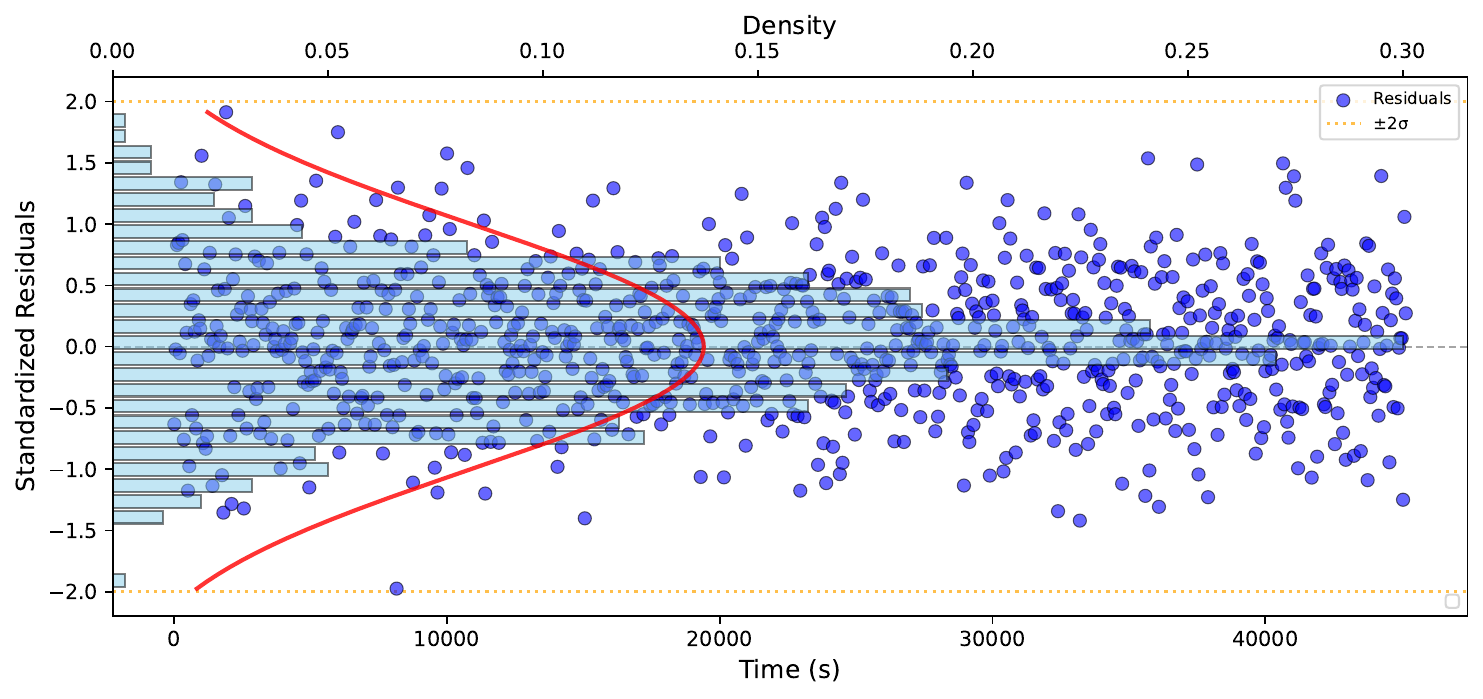}
  \end{minipage}
  \hfill
  \begin{minipage}[b]{0.48\linewidth}
    \centering
    \includegraphics[width=\linewidth]{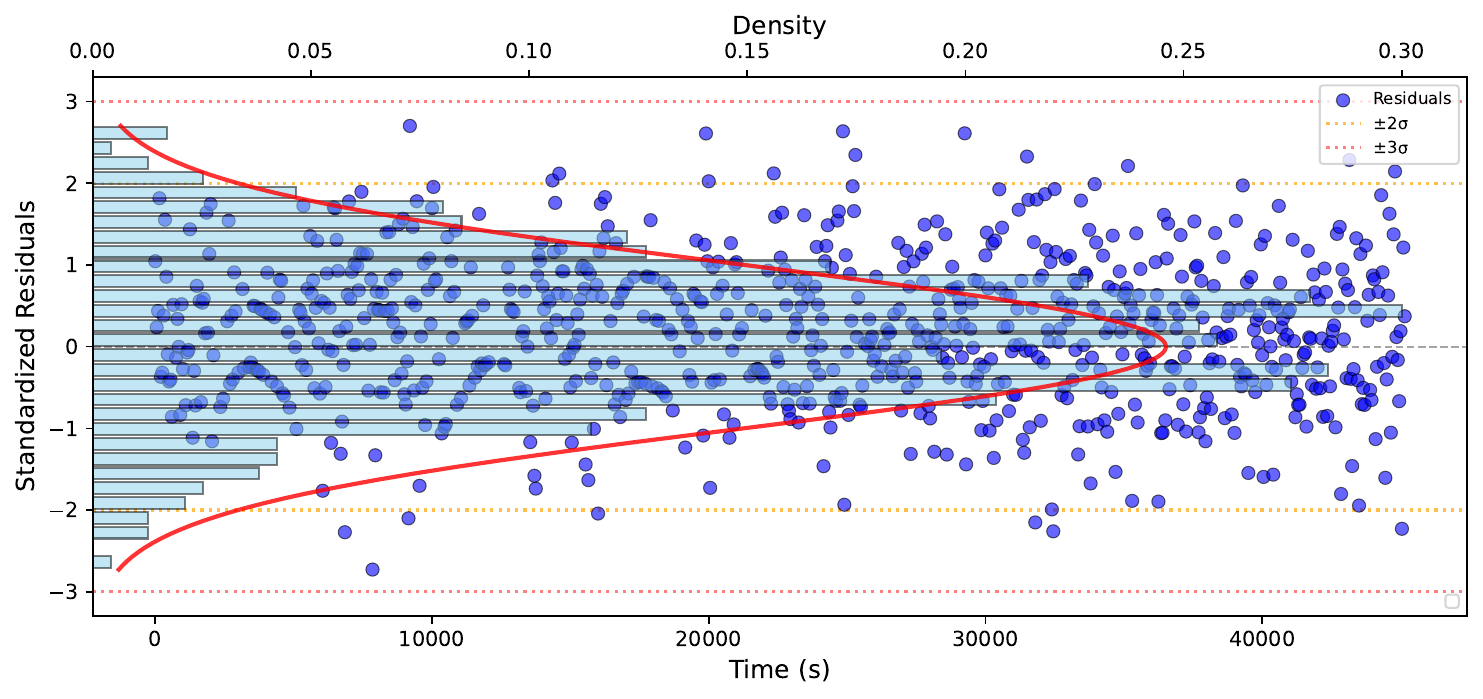}
  \end{minipage}

  \caption{
Gated-kernel analysis of PN\_0843270101.
The left column shows the 0.3--2.0 keV band.
The right column shows the 2.0--10.0 keV band.
The first row shows the light curves, the gated-kernel best fit, and the $1\sigma$ uncertainty.
The second row shows the inferred post-transition weight $q(t)$.
The third row shows the standardized residuals and their density distributions.
The vertical dashed lines mark the parametric gate centers.
  }
  \label{Fig2}
\end{figure*}

Figure~\ref{Fig2} presents the gated-kernel estimate of the transition time for PN\_0843270101. 
The left and right columns show the 0.3--2.0 keV and 2.0--10.0 keV bands, respectively. 
From top to bottom, the panels show the light curves with the gated-kernel best fit and $1\sigma$ uncertainty, the inferred post-transition weight $q(t)$, and the standardized-residual time series with the corresponding marginal density estimates. 
The vertical dashed lines mark the parametric gate centers, and the shaded regions indicate the effective 10--90 percent transition intervals.

The inferred gate behavior differs between the two bands. 
In the 0.3--2.0 keV band, $q(t)$ evolves gradually from the pre-transition covariance state to the post-transition covariance state, with a parametric center at $t_c\simeq11.3~{\rm ks}$ and an effective transition width of $\Delta t_{10\text{--}90}\sim40~{\rm ks}$. 
In the 2.0--10.0 keV band, the transition is much sharper: $q(t)$ remains close to zero before $\simeq23.5~{\rm ks}$ and then rapidly rises toward unity, with $t_c\simeq23.5~{\rm ks}$ and $\Delta t_{10\text{--}90}\simeq2.4~{\rm ks}$. 
The residual diagnostics improve relative to the single-state fits, indicating that the gated model captures much of the time-dependent residual structure seen in the stationary GP fits.
We have also analyzed the full-band light curve and obtained results consistent with those for the 0.3–2 keV band. This agreement is expected, since soft-band photons dominate the total counts.

% We emphasize that the gated-kernel fit is used here to estimate the transition time, not as a fully marginalized Bayesian change-point model. 
% The gated-kernel fit is used only to define candidate split times; the covariance-state change is then assessed using independent evidence comparisons in the resulting segments.
% To test whether the data on the two sides of the inferred transition favor different covariance states, we perform separate fixed-split evidence comparisons in the pre- and post-transition intervals.  
We next use the gate-derived candidate split times to test whether the
pre- and post-transition intervals support different covariance states.
For a given split time, we define
\begin{equation}
\Delta \ln Z_{\rm split}
=
\ln Z_{\rm pre}
+
\ln Z_{\rm post}
-
\ln Z_{\rm whole},
\end{equation}
where $\ln Z_{\rm pre}$ and $\ln Z_{\rm post}$ are the evidences of the preferred models in the two segments, and $\ln Z_{\rm whole}$ is the evidence of the preferred single-state model for the full exposure. 
This quantity measures the evidence gain of the fixed-split model relative to the full-exposure model.

In the 0.3--2.0 keV band, splitting the light curve at the parametric gate center, $t_c\simeq11.3~{\rm ks}$, does not provide a clean separation between the two stochastic covariance states. 
Before this split, the Mat\'ern-3/2 model is only marginally favored over the DRW model, with $\Delta \ln Z=0.9$; after the split, the DRW model is decisively favored, with $\Delta \ln Z=5.1$. 
However, the segmented model gives only $\Delta \ln Z_{\rm split}=0.4$ relative to the full-exposure single-state model. 
Thus, a split at the soft-band gate center is not strongly supported by the evidence.

This weak split preference is consistent with the broad and gradual nature of the soft-band gate. 
The parametric center marks the midpoint of the underlying logistic component, but, when the residual correction is included, it need not coincide exactly with the midpoint of the full fitted gate. 
Moreover, because the inferred 10--90 percent interval is very broad, the lower crossing lies partly outside the observed exposure. 
Thus, $t_c\simeq 11.3~{\rm ks}$ should not be interpreted as a clean boundary between two separated stochastic states.

As a diagnostic check, we therefore also test an effective split within the observed part of the broad transition. 
Specifically, we use the midpoint between the start of the observation and the inferred upper end of the visible transition region, giving $t_{\rm split}=17592~{\rm s}$ and $q(t)=0.68$. 
This split is not an independent change-point estimate, but a fixed boundary used to test whether the two sides of the broad soft-band transition have different preferred covariance states. 
With this split, the pre-transition interval decisively favors the Mat\'ern-3/2 model over the DRW model, with $\Delta \ln Z=8.6$, while the post-transition interval decisively favors the DRW model over the Mat\'ern-3/2 model, with $\Delta \ln Z=11.5$. 
The segmented model is also decisively favored over the full-exposure single-state model, with $\Delta \ln Z_{\rm split}=16.2$. 
Thus, although the soft-band gate center alone is not a useful segmentation boundary, the soft band supports a Mat\'ern-3/2-to-DRW change when the broad transition width is taken into account.

The 2.0--10.0 keV band provides a cleaner and more localized case. 
At the gate-defined hard-band split, the pre-transition interval significantly favors the Mat\'ern-3/2 model over the DRW model, with $\Delta \ln Z=2.3$, whereas the post-transition interval decisively favors the DRW model over the Mat\'ern-3/2 model, with $\Delta \ln Z=5.6$. 
The split-evidence comparison gives $\Delta \ln Z_{\rm split}=13.9$, providing decisive evidence for a two-state description in the hard band.

Collectively, the two bands support the same qualitative change in the dominant stochastic variability: the pre-transition variability is Mat\'ern-3/2-like, whereas the post-transition variability is DRW-like. 
The hard band localizes this switch sharply near $\sim 23.5~{\rm ks}$, while the soft band shows a much broader transition for which the gate center alone is not a clean segmentation boundary. 
The result therefore points to a broadband change in stochastic character, with different degrees of temporal localization in the two energy bands.
An influence from counting statistics cannot be completely excluded,
particularly given the lower hard-band count rate and correspondingly
larger statistical uncertainty. Nevertheless, the hard-band transition
remains sharply localized and the fixed-split model is strongly favored,
while both energy bands show the same Mat\'ern-3/2-to-DRW change.

\begin{figure}
  \includegraphics[width=\linewidth]{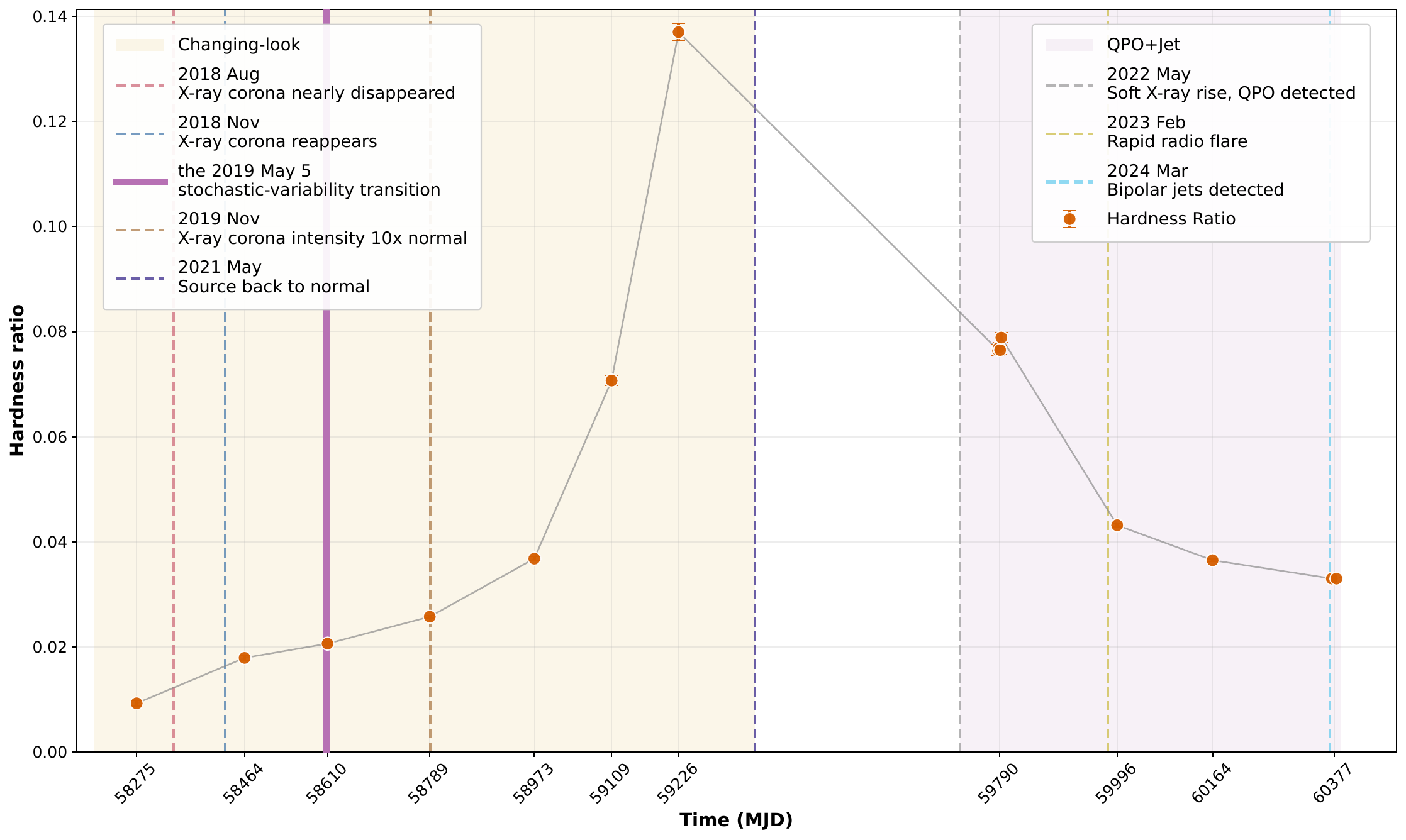}
  \caption{
Long-term evolution of the X-ray hardness ratio of 1ES 1927+654.
The hardness ratio is defined as the 2.0--10.0 keV count rate divided by the 0.3--2.0 keV count rate.
The orange points show the measured values, and the grey line connects them in time order.
The yellow shaded region marks the 2018--2021 changing-look recovery phase.
The purple shaded region marks the 2022--2024 QPO plus jet phase.
The vertical lines are described below in chronological order, from the beginning to the end of the dataset.
The rose dashed line marks the disappearance of the X-ray corona in 2018 August. 
The blue dashed line marks the reappearance of the X-ray corona in 2018 November. 
The thick solid line marks the 2019 May 5 stochastic-variability transition found in this work, where the preferred covariance changes from Matérn-3/2 to DRW. 
The brown dashed line marks the strong coronal brightening in 2019 November. 
The indigo dashed line marks the return to the pre-flare state in 2021 May. 
The grey dashed line marks the soft X-ray rise and QPO detection from 2022 May. 
The muted yellow dashed line marks the rapid radio flare in 2023 February. 
The cyan dashed line marks the bipolar jet detection in 2024 March.
  }
  \label{Fig3}
\end{figure}

Figure~\ref{Fig3} places the 2019 May 5 stochastic-variability transition in the broader X-ray evolution of 1ES 1927+654. 
The hardness ratio is defined as the 2.0--10.0 keV count rate divided by the 0.3--2.0 keV count rate. 
Following the phase classification of \citet{2025ApJ...981..125L}, we separate the evolution into the 2018--2021 changing-look recovery phase and the 2022--2024 QPO-plus-jet phase.

During the changing-look recovery phase, the hardness ratio rises from $\sim0.015$ in 2018 to a maximum of $\sim0.14$ in 2021. 
This hardening occurred after the near-disappearance of the X-ray corona in 2018 August and its reappearance in 2018 October--November \citep{2021ApJS..255....7R,2022ApJ...931....5L,2022ApJ...934...35M}. 
The source subsequently reached a much brighter coronal state in 2019 November, when the X-ray emission was reported to be about ten times stronger than its preflare level, before returning close to the preflare state by 2021 May.

The thick solid line marks the 2019 May 5 observation PN\_0843270101, in which we identify the stochastic-variability transition. 
Within this single exposure, the gated-kernel analysis indicates a broad soft-band transition with a parametric center at $t_c\simeq 11.3~{\rm ks}$ and a sharper hard-band transition at $t_c\simeq 23.5~{\rm ks}$. 
Across the transition, the preferred covariance changes from a Mat\'ern-3/2-like state to a DRW-like state. 
Thus, the timing-domain transition occurs after the corona had reappeared, but before the later pronounced hardening and brightening of the X-ray emission.

During the QPO-plus-jet phase, the hardness ratio decreases again, from $\sim0.08$ in 2022 to $\sim0.03$--$0.04$ in 2023--2024. 
This phase includes the soft X-ray rise and millihertz-QPO detection from 2022 May \citep{2025Natur.638..370M}, the rapid radio flare in 2023 February, and the spatially resolved bipolar jets detected in 2024 March \citep{2025ApJ...979L...2M}.

\subsection{Long-term evolution of stochastic variability states}
\label{sec:long_term_states}

\begin{figure} 
  \centering
  \includegraphics[width=\linewidth]{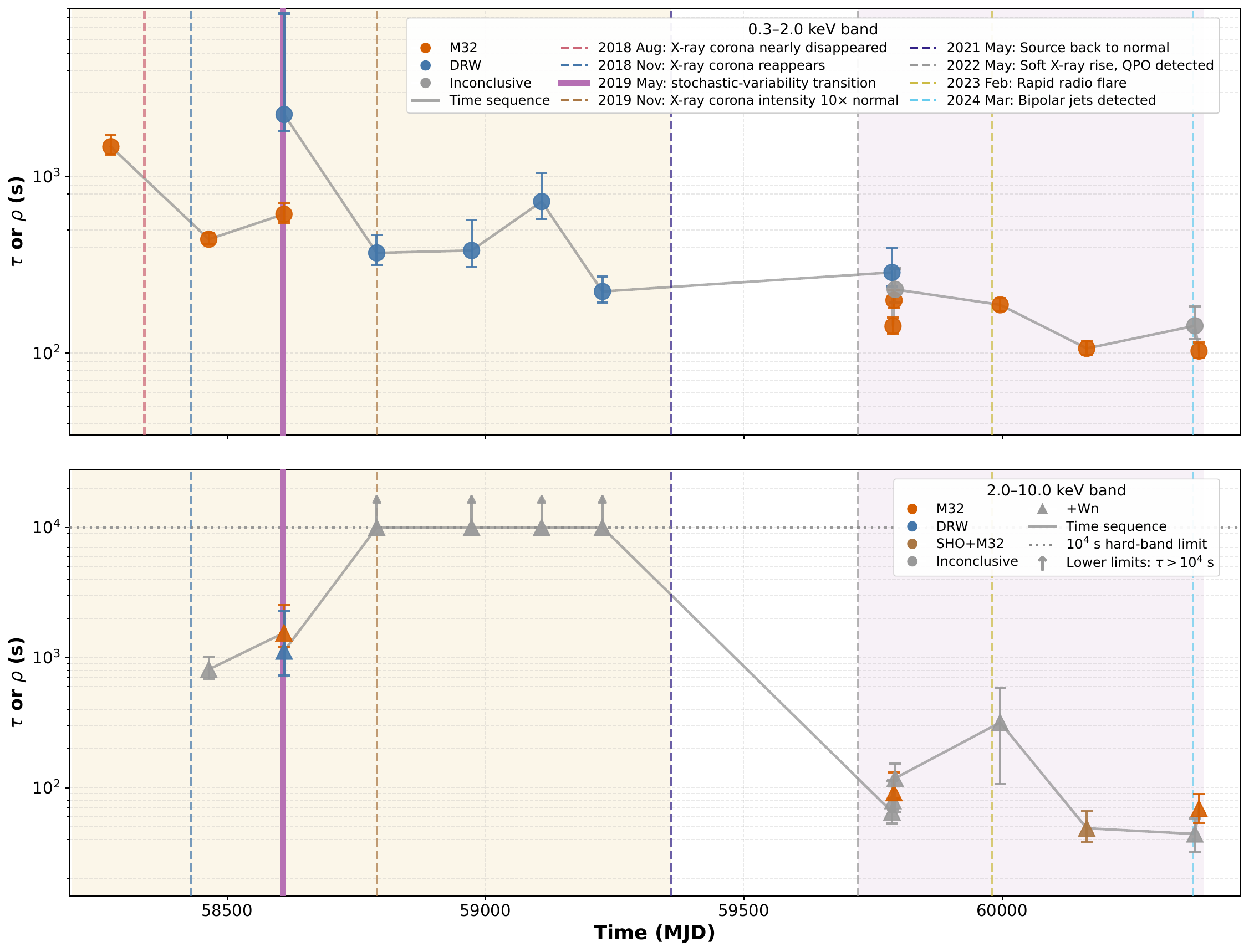}   
  \caption{
  Long-term evolution of the GP characteristic timescale in the 0.3--2.0 keV band and 2.0--10.0 keV band. Symbols indicate the preferred covariance model in each observation or analyzed sub-interval; arrows mark lower limits on the characteristic timescale in the hard band.
  }
  \label{Fig4}
\end{figure}

In the 0.3--2.0 keV band, Figure~\ref{Fig4} shows that the preferred covariance alternates between Mat\'ern-3/2 and DRW models over the 2018--2024 evolution.
The Mat\'ern-3/2 model is strongly or decisively preferred in eight cases, the DRW model in six cases, and two cases remain inconclusive. 
Thus, the soft-band stochastic variability does not remain in a single covariance state, but changes with time. 
The corresponding characteristic timescale is $\sim1500~{\rm s}$ in mid-2018, decreases to a few hundred seconds in late 2018, and remains mostly in the range of $\sim200$--$700~{\rm s}$ during 2019--2021. 
After 2022, it stays short, typically $\sim100$--$300~{\rm s}$, indicating that the soft-band variability remains rapid during the later QPO-plus-jet phase.

In the 2.0--10.0 keV band, we include an additional white-noise component to account for extra uncorrelated scatter, as the hard-band count rates are lower and the short-timescale variance is less well constrained. 
The model selection is therefore less decisive than in the soft band. 
The Mat\'ern-3/2 model is preferred in three observations, the DRW model in one observation, and an SHO plus Mat\'ern-3/2 model in one observation, while nine cases are inconclusive; four of these yield only lower limits above $10^4~{\rm s}$ for the characteristic timescale. 

Nevertheless, the inferred timescale is of order $\sim10^3~{\rm s}$ in late 2018 and early 2019, becomes longer during the 2019--2020 recovery period, and then decreases after 2021 to tens--hundreds of seconds by late 2022--early 2024. 
This evolution suggests a transition from a slower hard-band stochastic state during coronal recovery to faster variability in the later QPO-plus-jet phase.

\subsection{Later emergence of SHO-like QPO components}

During the 2022--2024 QPO+Jet phase, we fitted the 2.0--10.0 keV
light curves with SHO+DRW+Wn and SHO+Mat\'ern-3/2+Wn covariance models.
In these models, the SHO component captures the quasi-periodic
variability, the DRW or Mat\'ern-3/2 component represents the broadband
stochastic variability, and Wn accounts for additional uncorrelated
noise. Bayesian evidence was used to select the preferred covariance
model. The corresponding SHO parameters, $\nu_0$ and $Q$, were then used
to characterize the characteristic frequency and coherence of the
quasi-periodic component.

The $\nu_0$ increases from $\simeq 1.0\times10^{-3}~{\rm Hz}$ in late 2022 to $\simeq 1.6\times10^{-3}~{\rm Hz}$ around the 2023 radio flare, and then to $\simeq 2.3\times10^{-3}~{\rm Hz}$ in 2023--early 2024. 
This evolution is consistent with the reported upward drift of the millihertz QPO frequency in 1ES 1927+654 \citep{2025Natur.638..370M}. $Q$ increases with time.
It rises from $\simeq 10$ near the radio-flare epoch to $\simeq 13$ in 2023.
It then reaches $\simeq 60$ by early 2024.
This suggests that the SHO-like component becomes more coherent with time.

\section{Discussion}
\label{sec:scenario}

The GP analysis reveals two timing-domain signatures during the 2018--2024 evolution of 1ES 1927+654. 
The first is an X-ray stochastic-variability transition during the changing-look recovery phase. 
The second is the later emergence of SHO-like components associated with the known millihertz QPO. 
We discuss these results below.

\subsection{A stochastic-variability transition during coronal recovery}

The main result of this work is the identification of a change in the dominant stochastic variability during the recovery of the X-ray corona. 
In the 2019 May 5 observation PN\_0843270101, the gated-kernel analysis indicates a transition from a Mat\'ern-3/2-like covariance state to a DRW-like covariance state within a single continuous exposure. 
The hard band localizes this transition sharply near $t_c\simeq 23.5~{\rm ks}$, whereas the soft band shows the same qualitative change over a much broader interval. 
This timing-domain transition occurs after the X-ray corona had reappeared, but before the later pronounced hardening and brightening of the coronal emission.

This kernel transition should be interpreted phenomenologically. 
The Mat\'ern-3/2 and DRW kernels describe different stochastic characters of the X-ray light curves. 
A Mat\'ern-3/2-like process represents a comparatively smooth, finite-memory correlated variability pattern \citep{2006gpml.book.....R}, whereas a DRW-like process corresponds to a rougher, shorter-memory red-noise process \citep{2010ApJ...721.1014M}. 
% The observed Mat\'ern-3/2-to-DRW change therefore indicates that the dominant X-ray variability evolved from a smoother correlated state to a more stochastic, relaxation-dominated state.
The observed Matérn-3/2-to-DRW change therefore indicates that the dominant X-ray variability evolved from a smoother correlated state
to a rougher, shorter-memory state consistent with a first-order
stochastic relaxation process.

We define this covariance-state change as a \emph{stochastic-variability transition feature}, characterized by a transition center $t_c$, an effective width $\Delta t_{10\text{--}90}$, and distinct pre- and post-transition GP covariance states. 
To our knowledge, this is among the first uses of such a quantifiable covariance-transition descriptor in the X-ray timing study of a recovering AGN corona.

In the context of the known coronal collapse and recovery of 1ES 1927+654 \citep{2020ApJ...898L...1R, 2021ApJS..255....7R}, this transition may provide an early timing-domain signature of a reconfiguration of the inner disk--corona system. Its occurrence before the later increase in hardness ratio suggests that the stochastic variability changed before the spectral hardening became prominent. 
One possible interpretation is that the re-forming corona first changed its internal variability structure, followed by a more visible spectral evolution. 
Physical processes that could contribute to such a change include enhanced local magnetic heating, increased turbulent dissipation, or weakened coupling between the soft-photon field and the hard X-ray-emitting plasma. 
In this interpretation, the Mat\'ern-3/2-to-DRW transition traces increased stochasticity, or a loss of temporal coherence, in the recovering corona.

The long-term GP results further suggest that this transition was embedded in a broader evolution of the stochastic X-ray background. 
In the soft band, the preferred covariance alternates between Mat\'ern-3/2 and DRW states, while the characteristic timescale decreases from $\sim1500~{\rm s}$ in mid-2018 to $\sim100$--$300~{\rm s}$ after 2022. 
The hard band is less tightly constrained because of lower count rates and additional uncorrelated scatter, but its inferred timescale also evolves from longer values during the 2019--2020 recovery period to tens--hundreds of seconds after 2021. 
These trends suggest that the recovering corona did not simply return to a fixed preflare stochastic state; instead, its broadband variability evolved toward faster stochastic fluctuations before and during the later QPO-plus-jet phase.

\subsection{Later SHO-like components and the millihertz QPO}

The second timing signature appears in the later 2022--2024 observations, where the hard-band light curves show SHO-like covariance components. 
These components are phenomenologically associated with the known millihertz QPO in 1ES 1927+654 \citep{2025Natur.638..370M,2026arXiv260416688M}. 
Unlike the 2019 transition, which changes the broadband stochastic character of the light curve, the later SHO components describe a more coherent oscillatory contribution superposed on the stochastic X-ray variability.

As a simple interpretive picture, an oscillatory coronal mode may have a characteristic angular frequency
\begin{equation}
\omega_0 \approx \frac{\pi v_{\rm A}}{L},
\end{equation}
where $v_{\rm A}$ is the Alfv\'en speed and $L$ is the effective size of the oscillating region. 
In this picture, an increase in $\omega_0$ may reflect a decrease in the effective size of the oscillator, an increase in the relevant wave speed, or both. 
The quality factor can be written schematically as
\begin{equation}
Q=\frac{\omega_0}{2\gamma_{\rm total}},
\end{equation}
where $\gamma_{\rm total}$ represents the effective damping rate. 
The inferred SHO parameters show a systematic evolution: the characteristic frequency increases from $\simeq1.0~{\rm mHz}$ to $\simeq2.3~{\rm mHz}$, while $Q$ rises from $\simeq10$ to $\simeq60$. 
This suggests that the SHO-like component becomes both faster and more coherent during the QPO-plus-jet phase. 
The frequency increase is consistent with an oscillation moving toward shorter characteristic dynamical timescales, while the rise in $Q$ suggests weaker effective damping or improved phase coherence \citep{2026arXiv260401901D}.

The 2019 stochastic-variability transition and the later SHO-like QPO evolution indicate that the recovery of 1ES 1927+654 was not a simple monotonic return to its preflare state. 
Instead, the timing properties suggest a sequence of changes in the inner accretion flow and corona: first, a reorganization of the broadband stochastic variability during the changing-look recovery phase, and later, the development of a faster and more coherent oscillatory component during the QPO-plus-jet phase. 

\section{Summary and Conclusions}\label{sec:conclusions}
We have studied the stochastic X-ray variability of the changing-look AGN 1ES 1927+654 during its 2018--2024 evolution using XMM-Newton EPIC-pn light curves in the 0.3--2.0 keV and 2.0--10.0 keV bands. 
The light curves were modeled with GP covariance components including Mat\'ern-3/2, DRW, SHO, and additional white-noise terms, and Bayesian evidence was used to identify the preferred phenomenological variability state.

The main result is an X-ray stochastic-variability transition during the changing-look recovery phase. 
In the 2019 May 5 observation PN\_0843270101, the preferred covariance changes from a Mat\'ern-3/2-like state to a DRW-like state within a single continuous exposure. 
The transition is sharply localized in the hard band, with a gate center at $t_c\simeq 23.5~{\rm ks}$, while the soft band shows the same qualitative change over a broader interval. 
This timing-domain transition occurs after the X-ray corona had reappeared, but before the later pronounced hardening and brightening of the coronal emission. 
Phenomenologically, it indicates that the dominant X-ray variability evolved from a smoother, finite-memory correlated process to a rougher, shorter-memory red-noise process. 
We interpret this as an early timing-domain signature of disk--corona reconfiguration during coronal recovery.

In the later 2022--2024 observations, the hard-band light curves show SHO-like components phenomenologically associated with the known millihertz QPO. 
The characteristic frequency increases from $\simeq1.0~{\rm mHz}$ to $\simeq2.3~{\rm mHz}$, while $Q$ rises from $\simeq10$ to $\simeq60$, suggesting that the SHO-like component becomes faster and more coherent during the QPO-plus-jet phase. 
Collectively, the Mat\'ern-3/2-to-DRW transition and the later SHO evolution suggest that the recovery of 1ES 1927+654 proceeded from broadband stochastic reorganization to the development of a more coherent oscillatory component. 
GP-based time-domain inference therefore provides a useful phenomenological probe of stochastic-variability changes in recovering AGN coronae.

\begin{acknowledgments}
\centerline{Acknowledgments}
D.Y. acknowledges support from the National Natural Science Foundation of China (grant No.~12393852) and the Yunnan Provincial Science and Technology Department Foundation (grant No.~202601AT070175).
This work is based on observations obtained with \textit{XMM-Newton}, an ESA science mission with instruments and contributions directly funded by ESA Member States and NASA.
\end{acknowledgments}

\bibliography{d-j}{}
\bibliographystyle{aasjournal}	
\end{document}